\documentclass[10pt,twocolumn,a4paper]{article}
\usepackage[T1]{fontenc}
\usepackage[utf8]{inputenc}
\usepackage[a4paper,top=20mm,bottom=22mm,left=18mm,right=18mm,columnsep=6mm]{geometry}
\usepackage{amsmath,cite,url}
\usepackage{graphicx}
\usepackage{color}
\usepackage{multirow}
\usepackage{amssymb}
\usepackage[font=small,labelfont=bf]{caption}
\usepackage[bookmarks=true,hidelinks]{hyperref}

\title{Dance to Music Generation leveraging Pre-training with Unpaired data and Contrastive Alignment}
\author{%
  Ryota Kimura$^{1,2}$ \quad
  Sangheon Park$^{1,3}$ \quad
  Natalia Polouliakh$^1$ \quad
  Taketo Akama$^1$\\[0.5em]
  \small $^1$Sony Computer Science Laboratories \quad
  $^2$Keio University \quad
  $^3$Georgia Institute of Technology\\
  \small\texttt{ryokimu-2000@keio.jp, sangheon@gatech.edu, nata@sony.csl.co.jp, taketo.akama@gmail.com}
}
\date{}

\hypersetup{
  pdftitle={Dance to Music Generation leveraging Pre-training with Unpaired data and Contrastive Alignment},
  pdfauthor={Ryota Kimura, Sangheon Park, Natalia Polouliakh, Taketo Akama},
  pdfsubject={Preprint on dance-conditioned music generation},
  pdfkeywords={dance-to-music generation, contrastive learning, diffusion models, motion conditioning}
}

\begin{document}

\maketitle

\begin{abstract}
Dance-to-music generation is a promising task for applications such as choreography support and automatic accompaniment, where temporal coordination between body movement and sound is essential. In particular, using human joint positions as the motion representation is attractive because they explicitly capture body dynamics while being lightweight, privacy-preserving, and easy to integrate with motion capture and pose-estimation pipelines. A central challenge in this setting, however, is the scarcity of high-quality paired dance–music data, since collecting accurately synchronized pairs is costly and often constrained by copyright and performance rights. This makes it difficult to train end-to-end models solely from paired data.
To address this issue, we propose a dance-conditioned music generation framework that efficiently exploits both unpaired and paired data. Our method combines pretrained unimodal encoders for motion and music, beat-guided contrastive pretraining to align their feature spaces, and a ControlNet-style conditioning module on top of a pretrained text-to-audio diffusion model. Experiments on AIST++ demonstrate that the proposed techniques improve both dance–music alignment and audio quality, as confirmed by quantitative and qualitative evaluations. Compared to a state-of-the-art method, our approach achieves superior dance alignment performance and competitive audio quality. Code is available at \url{https://github.com/kmraven/AudioLDM-ControlNet}.
\end{abstract}

\section{Introduction}

Dance-to-music generation, which generates music from dance motion, is a useful task in settings where temporal consistency between body movement and sound is important, such as choreography support and automatic accompaniment. In particular, using joint positions as dance information has important significance. Joint positions are a representation that allows information related to body movement to be handled relatively explicitly while separating it from appearance-related factors contained in video, and thus they have advantages from the viewpoints of clarifying the conditioning target and improving controllability. In addition, because joint positions are lightweight and highly anonymous, they are easy to handle in practical applications and are also well suited for integration with motion capture, pose estimation, and various sensor-processing pipelines. Therefore, dance-to-music generation using joint positions has unique importance both methodologically and practically. For this problem setting, several prior studies have already been conducted, mainly focusing on architectural design based on autoregressive models or diffusion models\cite{textual_inversion,you2024momu,aggarwal2021dance2music,zhu2022quantized,han2024dance2midi,tan2023motion,zhang2024dance2musicdiffusion}.

On the other hand, a central challenge in this problem setting is the lack of high-quality paired dance--music data. Collecting datasets that accurately synchronize dance motion data, such as time series of joint positions, with music is costly. It requires securing performers, preparing recording environments, annotating the data, and verifying synchronization. In addition, copyright issues related to music and rights associated with performances and choreography pose major barriers. Furthermore, because dance covers a wide range of styles, it is not easy to construct a large-scale paired dataset that sufficiently includes rare styles and individualized bodily expressions. As a result, compared with single-modal data such as motion-only or music-only data, paired data are inherently likely to remain small in scale, making end-to-end learning that assumes sufficient paired data difficult.

Therefore, to train a high-performance model for this task, it is important to appropriately utilize unpaired motion data and music data, and also to efficiently learn cross-modal correspondences under limited paired data.

In related studies, some of these directions have already been explored. As an approach that utilizes pretrained music generation models, a textual inversion-style method has been proposed that injects dance-derived information into the text space of a text-to-music model\cite{textual_inversion}. In addition, methods have been reported that handle both motion-to-music and music-to-motion by aligning motion and music features into a shared latent space through contrastive learning, thereby promoting the learning of highly accurate long-term music--dance synchronization\cite{you2024momu}. However, these promising directions have not necessarily been integrated in the context of dance-to-music generation. First, to the best of our knowledge, no study has explicitly introduced, on the motion side, a model pretrained on single-modal data in dance-to-music generation. Moreover, a framework that combines feature alignment through contrastive learning with the use of a pretrained text-to-audio model does not appear in existing studies.

Based on these observations, this study aims to establish a framework for dance-conditioned music generation that efficiently utilizes both unpaired and paired data. 
Specifically, we leverage unimodally pretrained models and align the feature spaces of motion and music in advance through beat-guided contrastive learning. 
Experiments on AIST++\cite{ai_choreographer} show that these mechanisms improve dance--music alignment without substantially degrading sound quality. 
Compared with a prior state-of-the-art method in a closely related setting that also builds on a pretrained text-to-audio model while handling auxiliary conditions other than dance, the proposed method achieves superior dance-alignment performance while maintaining competitive sound quality, with better objective audio-quality scores despite slightly lower subjective sound-quality ratings. 
These results suggest that integrating unimodal pretraining, cross-modal alignment, and pretrained audio generation is an effective direction for dance-conditioned music generation.

\begin{figure*}
  \centering
  \includegraphics[width=1\linewidth]{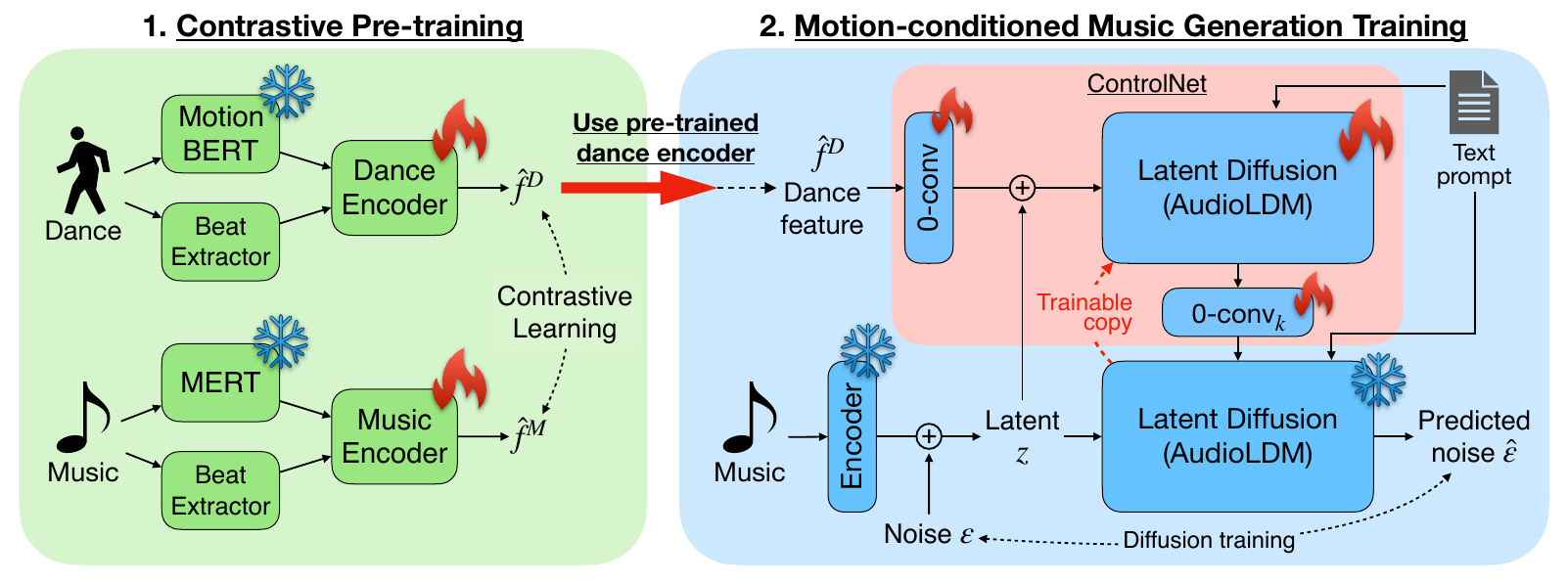}
\caption{Overview of the proposed two-stage dance-to-music generation framework. In the first stage, dance and music features are extracted using pretrained encoders with beat information and aligned through beat-guided contrastive learning. In the second stage, the pretrained dance encoder provides motion features that are projected and injected into a ControlNet-style adapter to condition a frozen AudioLDM backbone for motion-controlled music generation.}
  \label{fig:main}
\end{figure*}

\section{Related Work}\label{sec:related_work}

\subsection{Music Generation Models}
Prior work on music generation has mainly advanced the generative backbone itself, moving from autoregressive modeling to large-scale text-conditioned and diffusion-based systems. Jukebox modeled raw audio with a multi-scale VQ-VAE and autoregressive Transformers, showing that long-context music generation is possible directly in the waveform domain \cite{dhariwal2020jukebox}. Recent text-conditioned systems have shifted toward token- or codec-based generation, such as MusicLM, which formulates generation as a hierarchical sequence-to-sequence problem, and MusicGen, which uses a single-stage language model with efficient codebook interleaving for text- and melody-conditioned generation \cite{agostinelli2023musiclm,copet2023musicgen}. In parallel, diffusion-based methods such as Noise2Music and Mo\^usai generate music from text through cascaded or latent diffusion pipelines \cite{huang2023noise2music,schneider2023mousai}. AudioLDM further showed that latent diffusion can be trained in an audio latent space derived from CLAP, while Music ControlNet demonstrated that diffusion backbones can be extended with precise time-varying controls such as melody, dynamics, and rhythm \cite{audioldm,wu2024musiccontrolnet}.

\subsection{Dance-Conditioned Music Generation Models}
Prior work on dance-conditioned music generation has mainly explored how motion cues are represented and injected, including search-based systems, symbolic-generation pipelines, raw-audio synthesis, and adaptations of pretrained music generators. Aggarwal and Parikh presented early search-based offline generation and an online neural approach for dance-driven music generation \cite{aggarwal2021dance2music}. D2M-GAN conditions generation on dance video frames and human body motions and uses vector-quantized audio representations to synthesize complex dance music \cite{zhu2022quantized}. Dance2MIDI instead formulates the task in the symbolic domain, encoding dance motion with a graph convolutional network and generating multi-track MIDI with Transformer- and BERT-based modules \cite{han2024dance2midi}. For audio generation, Tan et al.  conditions a latent diffusion model on 3D motion and genre labels \cite{tan2023motion}, and Dance2Music-Diffusion uses SMPL-based motion features, a Transformer motion encoder, and latent diffusion for music synthesis \cite{zhang2024dance2musicdiffusion}. Particularly close to keypoint-centered conditioning, Li et al. derive rhythm from 2D joint positions and map rhythm and genre information into the textual space of pretrained text-to-music models through encoder-based textual inversion \cite{textual_inversion}.

\subsection{Contrastive Pretraining of Condition Encoders for Conditional Generation}
A separate line of work studies contrastively aligned encoders as reusable condition interfaces for generation, learning cross-modal representations before training the generator. In the image domain, CLIP learns a joint image--text embedding space by contrastive pretraining \cite{radford2021learning}, and hierarchical generation with CLIP latents conditions a diffusion decoder on CLIP image embeddings produced from text \cite{ramesh2022hierarchical}. In audio, AudioLDM uses CLAP-derived latent representations to train latent diffusion while using text embeddings for conditioning at sampling time \cite{audioldm}. For music--dance modeling, BeatDance learns beat-aware aligned music and dance representations with contrastive learning for retrieval \cite{beatdance}, while MoMu-Diffusion introduces a bidirectional contrastive rhythmic VAE to obtain modality-aligned motion and music latents before training a Transformer-based diffusion model for cross-modal and joint generation \cite{you2024momu}. These studies establish the usefulness of contrastively aligned condition spaces, while in dance-conditioned music generation, contrastive alignment and pretrained text-to-audio backbones have largely been explored in separate lines of work.

\section{Proposed Framework}\label{sec:methodology}
Our framework decouples dance-to-music generation into two phases: (1) \textbf{Cross-modal Representation Learning}, which aligns motion and music in a shared latent space via contrastive pretraining, and (2) \textbf{Conditional Generation}, where a ControlNet-style adapter steers a pretrained AudioLDM backbone using the aligned features. 
Specifically, Section \ref{subsec:feature_extraction} outlines the feature extraction process for both motion and music modalities.
Section \ref{subsec:contrastive_pretraining} introduces our beat-guided contrastive pretraining approach to align these extracted representations.
Finally, Section \ref{subsec:generation} details how we inject the aligned motion features into the pretrained audio generation backbone using a ControlNet-style adapter.

\subsection{Multimodal Feature Encoding}\label{subsec:feature_extraction}
To exploit the structural priors of human movement and acoustic semantics, we utilize pretrained feature extractors trained on large-scale unimodal data.
Our feature extraction design is based on \cite{beatdance} so as to align with the subsequent contrastive learning strategy, in that beat features are additionally incorporated as auxiliary cues alongside direct features extracted from each modality.

\subsubsection{Motion Encoding}
Let $d=\{ \mathbf{p}_t \}_{t=1}^{T}$ denote the input 2D keypoint sequence, where $\mathbf{p}_t\in\mathbb{R}^{J\times 2}$ contains $J$ joints at frame $t$.
We first temporally resample the input sequence $d$ to a fixed number of frames $L{=}128$. We then feed this sequence into the pretrained \textit{MotionBERT} encoder ~\cite{motionbert} to extract the frame-wise motion features $f^D = \Gamma_{D}(d) \in\mathbb{R}^{L\times D_D}$, where $\Gamma_{D}(\cdot)$ denotes the frozen motion feature extractor and $D_D$ is the motion feature dimension.

To provide explicit rhythmic anchors, we additionally extract kinematic beat features from the motion sequence.
Specifically, we first obtain a binary kinematic beat sequence at the original frame rate by smoothing the frame-wise velocity
$v_t = \lVert \mathbf{c}_t - \mathbf{c}_{t-1} \rVert_2$
with a 1D Gaussian filter and marking local minima, where $\mathbf{c}_t$ is the root-centered joint mean at frame $t$.
The extracted beat sequence is then temporally downsampled to length
$L \cdot D_{BD}$ and reshaped into a matrix
$f^{BD} \in \{0,1\}^{L \times D_{BD}}$,
where $D_{BD}$ is an integer factor.
This representation allows the beat feature to be concatenated with the semantic motion features along the feature dimension.

\subsubsection{Music Encoding}
Given music waveform $m$, we first extract an acoustic embedding sequence from $m$ using \textit{MERT}~\cite{mert}.
Since the output sequence length of \textit{MERT} is generally different from $L$,
we temporally average the extracted sequence to obtain $L$ aligned music features $f^{M}=\Gamma_{M}(m)\in\mathbb{R}^{L\times D_M}$,
where $\Gamma_{M}(\cdot)$ denotes the frozen music feature extractor together with the temporal averaging operation, and $D_M$ is the music feature dimension.

We similarly extract music beats using Librosa~\cite{librosa}.
Specifically, we first detect beat timestamps from the paired music signal and convert them into a binary beat sequence at the original temporal resolution.
The resulting sequence is then temporally downsampled to length $L \cdot D_{BM}$ and reshaped into
$f^{BM} \in \{0,1\}^{L \times D_{BM}}$
where $D_{BM} \in \mathbb{N}$ is an integer factor.

\subsection{Beat-Guided Contrastive Pretraining}\label{subsec:contrastive_pretraining}
To establish a principled interface between modalities, we perform contrastive pretraining on paired dance–music data.

\subsubsection{Multimodal Fusion Strategy}
We adopt the dual-path transformer from \cite{beatdance}, augmented with an explicit Positional Encoding (PE) layer~\cite{vaswani2017attention} on the beat stream to better capture temporal order in short (5.12\,s) clips.
Let $\hat{f}^{M}, \hat{f}^{D}\in\mathbb{R}^{L\times D}$ denote the music and dance embeddings, where $L$ is the temporal length, and $D$ is the embedding dimension.
The fused representations are computed as:
\begin{equation}
\hat{f}^D = \sigma \left( W_f^D [ (f^D + f^{BD}) \oplus (f^D \odot f^{BD}) ] + b_f^D \right)
\end{equation}
and
\begin{equation}
\hat{f}^M = \sigma \left( W_f^M [ (f^M + f^{BM}) \oplus (f^M \odot f^{BM}) ] + b_f^M \right),
\end{equation}
where $W_f$ and $b_f$ are learnable parameters, followed by a Multi-head Attention block\cite{vaswani2017attention} where attended beats serve as queries over the semantic features.

\subsubsection{Temporal Maximization and Similarity Matrix}
Rather than applying Global Average Pooling as in \cite{beatdance}, we compute the segment-level similarity from batched, frame-wise $\ell_2$-normalized embeddings $\hat{\mathbf{f}}^{M}, \hat{\mathbf{f}}^{D}\in\mathbb{R}^{B\times L\times D}$, where $B$ is the batch size. Specifically, the similarity between the $i$-th music segment and the $j$-th motion segment in the batch is defined as the maximum frame-wise dot product:
\begin{equation}
s_{i,j}=\max_{1\le l\le L}\; \hat{\mathbf{f}}^{M}_{i,l}\cdot \hat{\mathbf{f}}^{D}_{j,l},
\end{equation}
where $\hat{\mathbf{f}}^{M}_{i,l}, \hat{\mathbf{f}}^{D}_{j,l}\in\mathbb{R}^{D}$ denote the $l$-th frame embedding of the $i$-th music segment and the $j$-th motion segment, respectively. This similarity is then used to drive the contrastive loss $\mathcal{L}_{\mathrm{NCE}}$ through the most salient motion--music alignments, preventing rhythmic signal dilution over long windows in low-resource settings.

\subsection{Motion-Conditioned Generation via ControlNet}\label{subsec:generation}
We build our conditional generator by augmenting a pretrained AudioLDM backbone with a ControlNet-style adapter.
AudioLDM is a latent diffusion model (LDM) that performs denoising in a compact audio latent space obtained from a pretrained VAE\cite{audioldm}.

\subsubsection{ControlNet Adapter for Motion Conditioning}
Given a music waveform $m$, the VAE encoder transforms it into a latent representation
$z_0 \in \mathbb{R}^{C \times F \times L}$, where $C$ and $F$ denote the channel and frequency dimensions of the latent space, respectively.
The diffusion process then perturbs $z_0$ with Gaussian noise to obtain $z_t$ at timestep $t$.

To condition generation on motion without overwriting the pretrained generative prior of AudioLDM, we follow the ControlNet design principle~\cite{controlnet}: we freeze the original denoising U-Net and attach a trainable conditional branch.
The conditioning signal $\hat{f}^{D}\in \mathbb{R}^{L \times D}$ is obtained from the dance input using our contrastively pretrained dance encoder, where $L$ is aligned with the latent temporal resolution of AudioLDM.
To interface with the AudioLDM backbone, the dance representation is projected to the input feature space of AudioLDM:
\begin{equation}
c = P(\hat{f}^{D}),
\end{equation}
where $P(\cdot)$ denotes a learnable linear projection that maps the embedding dimension $D$ to the feature dimensions used by AudioLDM, including channel and frequency.

The projected motion condition is then injected into the ControlNet input through a zero-initialized convolution:
\begin{equation}
x_t^{(\mathrm{ctrl})} = z_t + \mathrm{ZeroConv}(c),
\end{equation}
The conditional branch predicts a condition-dependent residual feature at each controlled block $k$:
\begin{equation}
\Delta h_k = \mathcal{F}_k(x_t^{(\mathrm{ctrl})}, t),
\end{equation}
where $\mathcal{F}_k(\cdot)$ denotes the $k$-th trainable ControlNet block.
The predicted residual is injected into the frozen backbone through another zero-initialized convolution:
\begin{equation}
h_k = h_k^{(\mathrm{base})} + \mathrm{ZeroConv}_k(\Delta h_k),
\end{equation}
where $h_k^{(\mathrm{base})}$ is the activation of the frozen backbone at block $k$.
Because all zero-convolution layers are initialized to zero, the overall network is identical to the original pretrained model at the beginning of training, enabling stable adaptation under limited paired data~\cite{controlnet}.

\subsubsection{Training Objective}
We train the conditional generator using the standard diffusion noise-prediction objective in AudioLDM's latent space \cite{audioldm}.
For a clean latent $z_0$, noise $\epsilon \sim \mathcal{N}(0, I)$, and timestep $t$, we construct $z_t$ with the forward noising process and optimize
\begin{equation}
\mathcal{L}_{\text{diff}} =
\mathbb{E}_{z_0, \epsilon, t}\left[
\left\|
\epsilon - \epsilon_{\theta,\theta_c}(z_t, t, \text{text}, \tilde{c})
\right\|_2^2
\right],
\end{equation}
where $\epsilon_{\theta,\theta_c}$ denotes the combined denoiser with a frozen AudioLDM backbone and a trainable ControlNet branch \cite{controlnet}.
Following classifier-free guidance practice, we randomly drop the text prompt during training, while keeping motion conditioning active, so that the model can generate with or without text at inference time \cite{audioldm}.

\subsubsection{Inference}
At inference, we sample an initial Gaussian latent $z_T \sim \mathcal{N}(0,I)$ and iteratively denoise it using DDIM sampling conditioned on the motion representation $c$ (and optional text).
The final latent is decoded by the VAE decoder to obtain the waveform \cite{audioldm}.

\section{Experiments}\label{sec:experiments}
\subsection{Dataset \& Setup}
\subsubsection{Dataset}
We conduct experiments on the AIST++ dataset\cite{ai_choreographer}, which provides paired dance and music data for dance-motion and music-related tasks.
Following prior work on dance-to-music generation, we use the same train/validation/test split protocol as a previous study\cite{textual_inversion} to ensure a fair comparison and avoid differences caused by data partitioning.
The resulting split contains 2,744 / 36 / 36 dance--music pairs for train / validation / test, respectively.

\subsubsection{Preprocessing}
For dance input, we use 2D joint positions.
The original motion sequences are provided at 60 fps, and we resample them to 25 fps before training.
Following MotionBERT-style preprocessing, we normalize joint positions in camera coordinates, 
and apply random scaling augmentation to the dance input during the training phase.

For music, the generation target is waveform audio.
We follow the AudioLDM preprocessing pipeline: audio is resampled to 16 kHz and encoded into latent features using the pretrained VAE (latent frame rate of 25 Hz).
At inference time, generated latents are decoded back into waveform audio.

In addition to dance and audio, we use text prompts as an auxiliary condition.
To construct the prompts, we use the top-50 tags from the MTG-Jamendo dataset~\cite{bogdanov2019mtg} and compute the cosine similarity between each music clip and each tag in the CLAP embedding space~\cite{clap}.
We then select the top-1 genre tag, top-3 instrument tags, and top-1 mood/theme tag, and instantiate the template
``A \{mood\} \{genre\} track featuring \{inst1\}, \{inst2\} and \{inst3\}.''

During conditional generation training, we apply synchronized time-stretch augmentation to the paired dance and music inputs with a probability of 70\%. 
The stretch factor is randomly sampled in the range of $[0.8, 1.0]$. 
We also apply classifier-free style text dropout following ControlNet practice, where the text condition is replaced with an empty string with a probability of 50\%.

\subsubsection{Training Details}
Our training consists of two stages: (i) contrastive pretraining of the dance encoder using paired dance--music data, and (ii) conditional diffusion training with the pretrained dance encoder as the conditioning feature extractor.

The dance encoder is pretrained using a contrastive objective on paired dance--music data. Specifically, the model is optimized for 150 epochs using the AdamW optimizer with a learning rate of $1.0 \times 10^{-6}$, a batch size of 32, and a weight decay of 0.2. During this stage, we project the modalities into a shared embedding space with a dimensionality of 256.

For conditional diffusion training, we train the model for 300k steps until the training loss sufficiently converges. 
We use Adam with a learning rate of $1.0\times10^{-4}$ and a batch size of 8. At inference time, we use DDIM sampling with 200 steps and a guidance scale of 3.5.

\begin{table*}[htbp]
\centering
\small
\setlength{\tabcolsep}{4pt}
\renewcommand{\arraystretch}{1.15}
\resizebox{\textwidth}{!}{%
\begin{tabular}{l|cccc|cc|ccc}
\hline
\multirow{3}{*}
& \multicolumn{6}{c|}{Dance--music Alignment}
& \multicolumn{3}{c}{\multirow{2}{*}{Sound Quality}} \\
\cline{2-7}
& \multicolumn{4}{c|}{vs. Ground-truth Music}
& \multicolumn{2}{c|}{vs. Ground-truth Dance}
& \multicolumn{3}{c}{} \\
\cline{2-10}
& BHS $\uparrow$
& BCS $\uparrow$
& F1 $\uparrow$
& TD $\downarrow$
& BAS $\uparrow$
& \begin{tabular}{c}MOS $\uparrow$\\(Dance alignment)\end{tabular}
& CLAP $\uparrow$
& FAD $\downarrow$
& \begin{tabular}{c}MOS $\uparrow$\\(Sound quality)\end{tabular} \\
\hline
Ground-truth Music
& -- & -- & -- & --
& 0.268 & 4.07
& -- & -- & 3.80 \\
\hline

\textbf{Ours}
& 0.535 & 0.510 & 0.386 & \underline{20.7}
& \underline{0.234} & \textbf{2.82}
& 0.681 & \textbf{4.56} & 2.65 \\

Ours (w/o Contrastive Pretraining)
& 0.546 & \textbf{0.536} & 0.415 & 24.5
& \textbf{0.238} & 2.39
& \underline{0.686} & \underline{4.86} & 2.35 \\

Ours (w/o MotionBERT)
& 0.511 & \underline{0.535} & \underline{0.431} & 26.3
& 0.217 & 2.60
& \textbf{0.703} & 5.44 & \underline{2.83} \\

AudioLDM (default)
& \underline{0.593} & 0.397 & 0.399 & 24.8
& 0.217 & 1.60
& 0.586 & 6.05 & 1.40 \\

Li et al.~\cite{textual_inversion}
& \textbf{0.655} & 0.433 & \textbf{0.467} & \textbf{13.7}
& 0.194 & \underline{2.66}
& 0.403 & 6.90 & \textbf{2.88} \\
\hline
\end{tabular}%
}
\caption{Quantitative evaluation and user study results on AIST++. Metrics are grouped into dance--music alignment and sound quality. The best and second-best results among generation methods are shown in \textbf{bold} and \underline{underlined}, respectively.}
\label{tab:results}
\end{table*}

\subsection{Evaluation}
We assess the proposed method from both objective and subjective perspectives.
For objective evaluation, we measure (1) fidelity to the reference music, in terms of rhythm and audio/semantic similarity, and (2) rhythmic alignment between the generated music and the input dance.
For subjective evaluation, we conduct a listener study that rates perceptual dance alignment and overall sound quality.

\smallskip
\noindent\textbf{Rhythmic Similarity.}
To measure rhythmic fidelity to the reference music, we adopt Beat Hit Score (BHS), Beat Coverage Score (BCS), F1, and Tempo Difference (TD) following \cite{textual_inversion}.
BHS and BCS measure the recall and precision of generated beats against reference beats within a 0.2-second tolerance window, and F1 is their harmonic mean.
TD reports the absolute tempo difference in BPM between generated and reference music.
Higher BCS, BHS, and F1 indicate better rhythmic agreement with the reference music, whereas lower TD indicates smaller tempo deviation from the reference.

\smallskip
\noindent\textbf{Audio / Semantic Similarity.}
To evaluate audio/semantic fidelity to the reference music, we report Fr\'{e}chet Audio Distance (FAD)~\cite{fad} and CLAP similarity~\cite{clap}.
FAD measures distributional similarity between generated and reference audio using VGGish embeddings, whereas CLAP similarity computes cosine similarity between generated and reference audio in the CLAP embedding space.
Lower FAD indicates higher audio fidelity and closer similarity to the reference, while higher CLAP similarity indicates that the generated audio is semantically closer to the reference.

\smallskip
\noindent\textbf{Dance--Music Alignment.}
To quantify rhythmic alignment between the input dance and the generated music, we adopt Beat Alignment Score (BAS) from \cite{ai_choreographer}:
\begin{equation}
\mathrm{BAS} = \frac{1}{m}\sum_{i=1}^{m}\exp\!\left(
-\frac{\min_{t_j^y\in B^y}\left\|t_i^x-t_j^y\right\|^2}{2\sigma^2}
\right),
\end{equation}
where $B^x=\{t_i^x\}$ and $B^y=\{t_j^y\}$ are kinematic and music beat times, respectively, and $\sigma$ is a normalization parameter following \cite{ai_choreographer}.
Higher BAS indicates better dance--music beat alignment.

\smallskip
\noindent\textbf{Subjective Evaluation.}
We conduct a Mean Opinion Score (MOS) study with 35 participants via a web-based survey.
Each participant rates 10 randomly sampled test examples on a 0--5 scale for
(1) \textit{Dance alignment}, which evaluates how well the generated music matches the rhythm and motion of the input dance, and
(2) \textit{Sound quality}, which evaluates the overall perceptual audio quality.

Taken together, these objective and subjective metrics provide a comprehensive evaluation of reference-music fidelity, dance-conditioning controllability, and perceptual audio quality.

\section{Results and Analysis}

We analyze the contribution of each proposed component and compare with prior work in Table~\ref{tab:results}. Our full model is compared against three ablations: (i) \textbf{w/o Contrastive Pretraining}, where the dance encoder is trained only during conditional diffusion training (extended to 400k steps to compensate for the absence of pretraining); (ii) \textbf{w/o MotionBERT}, where the MotionBERT-based motion representation is replaced with a simpler alternative; and (iii) \textbf{AudioLDM (default)}, which removes dance conditioning entirely.
As a prior-work baseline, we use Li et al.~\cite{textual_inversion}, since it addresses a closely related problem setting by building on a pretrained text-to-audio model and incorporating auxiliary input information beyond joint positions (genre labels in their method and text prompts in ours). 

\subsection{Ablation Study}
The ablation results show that the full model achieves the best MOS for dance alignment (2.82), outperforming all reduced variants. This indicates that the proposed components contribute jointly to perceptually better dance--music matching. In particular, removing contrastive pretraining drops sharply in dance-alignment MOS from 2.82 to 2.39, even though this variant achieving comparable or slightly better scores on most objective metrics (BCS, F1, FAD, CLAP, BAS). This dissociation between objective and perceptual scores suggests that contrastive pretraining instills a richer motion representation that manifests in perceptually meaningful dance--music correspondence, beyond what beat-level alignment metrics can capture. Similarly, removing MotionBERT lowers both BAS (0.217) and dance-alignment MOS (2.60), supporting the role of leveraging the feature extractors pretrained on large-scale unimodal data in providing more reliable conditioning features. Finally, the unconditional AudioLDM baseline performs substantially worse in BAS and dance-alignment MOS, confirming that explicit dance conditioning is essential. 

On the other hand, the variant without MotionBERT yields higher CLAP score and sound-quality MOS than the full model, suggesting that the MotionBERT-based representation may improve motion controllability while sacrificing some degree of audio quality and semantic fidelity. This trade-off may arise because, without pretrained MotionBERT features, the model has difficulty extracting informative motion cues and therefore tends to rely more on non-motion information. Alternatively, although MotionBERT provides a strong motion prior, it may suppress or discard musically relevant information implicitly contained in the dance signal, which could negatively affect semantic consistency and perceived sound quality.

\subsection{Comparison with Prior Work}

Table~\ref{tab:results} also compares the proposed method with the previously reported method of Li et al.~\cite{textual_inversion}, using its publicly available MusicGen-backbone variant. 
Under the same dataset split and evaluation protocol, our method improves BAS from 0.194 to 0.234 and dance-alignment MOS from 2.66 to 2.82, indicating that it better aligns the generated music with the input dance motion from the perspective of dance-conditioned generation. 
It should be noted that Li et al. achieve higher BHS, F1, and TD scores, suggesting stronger beat-level similarity to the reference music. 
However, these metrics compare the generated music with a single reference track, whereas multiple valid musical realizations can correspond to the same dance sequence in dance-to-music generation. 
Under this view, BAS and dance-alignment MOS provide more direct evidence of dance-conditioned alignment with the input motion, whereas BHS, BCS, F1, and TD should be interpreted as complementary measures of rhythmic similarity to the reference music.

Regarding sound quality, our method obtains a slightly lower MOS than Li et al., indicating that it does not fully match the subjective audio quality of the MusicGen-based baseline. However, our method substantially improves the objective sound-quality metrics, reducing FAD from 6.90 to 4.56 and increasing CLAP similarity from 0.403 to 0.681. These results suggest that, although our method is slightly behind the state-of-the-art baseline in perceived sound quality, it achieves competitive sound quality overall and even outperforms the baseline in objective evaluation.

This comparison should also be interpreted in light of the different generation backbones used in the two methods: Li et al. use MusicGen, whereas our framework is built on AudioLDM. 
Since the focus of this work is the proposed dance-conditioning mechanism rather than the audio generator itself, the ablation results in Table~\ref{tab:results} are also informative: compared with the AudioLDM (default) baseline, the proposed conditioning mechanism improves dance alignment without degrading sound-quality metrics. 
Thus, the sound-quality gap in subjective MOS is likely influenced by the underlying backbone, while the proposed conditioning mechanism itself remains compatible with competitive audio generation.

\section{Conclusion}\label{sec:conclusion}
In this work, we presented a dance-to-music generation framework that makes effective use of the information available for this task by combining pretrained unimodal encoders, beat-guided contrastive alignment, and a ControlNet-style conditioning module on top of a pretrained AudioLDM backbone. Using 2D joint positions as the motion representation, the proposed method incorporates motion structure, beat cues, music-side acoustic knowledge, and pretrained audio generation priors within a unified training framework for dance-conditioned music generation. Experiments on AIST++ showed improved dance–music alignment over relevant baselines, including gains in BAS and dance-alignment MOS relative to Li et al., while the ablation study supported the contribution of both contrastive pretraining and MotionBERT-based motion features. In terms of sound quality, although the proposed method obtained a slightly lower subjective MOS than Li et al., its better FAD and CLAP scores suggest that it remains competitive from the perspective of objective audio-quality evaluation. These results suggest that explicitly integrating the available unimodal and cross-modal information is an effective direction for dance-conditioned music generation.

\section{AI Usage Statement}
The authors used large language models to assist with the writing of the manuscript, excluding figure creation. Specifically, LLMs were used for drafting, editing, reorganizing, and improving the clarity of the text. All content, including claims, interpretations, and references, was reviewed, verified, and revised by the authors, who take full responsibility for the correctness, originality, and integrity of the manuscript.

\bibliographystyle{IEEEtran}
\bibliography{references}

\end{document}